\documentclass[%
 aip,
 jmp,%
 amsmath,amssymb,
 reprint,%
 nofootinbib,
floatfix,
]{revtex4-1}

\usepackage{amsmath,amssymb,bm,bbm,mathrsfs,amscd}
\usepackage{dsfont}
\usepackage{hyperref}
\usepackage{comment}
\usepackage{graphicx}
\usepackage{physics}
\usepackage{epstopdf}
\usepackage{epsfig}
\usepackage{tikz}

\usepackage{natbib}
\usepackage{appendix}
\usepackage{float}

\usepackage{caption}
\usepackage{subcaption}
\usepackage[normalem]{ulem}

\usepackage[T1]{fontenc}
\usepackage{mathpazo}

\usepackage{enumerate,pgfplots}
\usepackage{algpseudocode}
\pgfplotsset{ 
  compat=newest, 
  legend style =
  {font=\footnotesize \sffamily},
  label style = {font=\small\sffamily},
every tick label/.append style={font=\small}}
\usetikzlibrary {positioning}
\usepgfplotslibrary{fillbetween}
\usepackage{amsthm}
\usepackage{nomencl}
\usepackage{braket}
\usepackage{cancel}
\usepackage{mathbbol}
\usepackage{enumitem}
\usepackage{appendix}
\usepackage{placeins}
\usepackage{multirow}

\usepackage{float}
\usepackage[all]{nowidow}
\usepackage{mhchem}
\usepackage[normalem]{ulem}

\newcommand{\eqnref}[1]{Eq.~\eqref{#1}}
\newcommand{\figref}[1]{Fig.~\ref{#1}}

\newcommand{\tabref}[1]{Table~\ref{#1}}

\newcommand{\cminv}{cm${}^{-1}$}

\newcommand*{\citen}[1]{%
  \begingroup
    \romannumeral-`\x % remove space at the beginning of \setcitestyle
    \setcitestyle{numbers}%
    \cite{#1}%
  \endgroup   
}

\newcommand{\etal}{~\emph{et al.}}

\DeclareSymbolFontAlphabet{\mathbbm}{bbold}
\DeclareSymbolFontAlphabet{\mathbb}{AMSb}

\DeclareSymbolFont{newfont}{OML}{cmm}{m}{it}% Computer Modern math font
\DeclareMathSymbol{\Epsilon}{3}{newfont}{15}% Symbol 15
\DeclareMathSymbol{\Varrho}{3}{newfont}{37}% Symbol 37

\usepackage{graphicx} % Required for inserting images

\date{\today}

\begin{document}

\title{Anharmonic phonons via quantum thermal bath simulations}
\author{T. Baird}
 \affiliation{Centre Europ\'een de Calcul Atomique et Mol\'eculaire (CECAM), Ecole Polytechnique F\'ed\'erale de Lausanne, 1015 Lausanne, Switzerland}
\author{R. Vuilleumier}%
\affiliation{ 
PASTEUR, D\'epartement de chimie, \'Ecole normale sup\'erieure, PSL University, Sorbonne Universit\'e. CNRS, 75005 Paris, France%\\This line break forced with \textbackslash\textbackslash
}%
\author{S. Bonella}
 \email{sara.bonella@epfl.ch}
 \affiliation{Centre Europ\'een de Calcul Atomique et Mol\'eculaire (CECAM), Ecole Polytechnique F\'ed\'erale de Lausanne, 1015 Lausanne, Switzerland}

\begin{abstract}
Lattice vibrations within crystalline solids, or phonons, provide information on a variety of important material characteristics, from thermal qualities to optical properties and phase transition behaviour.   When the material contains light ions, or is subjected to sufficiently low temperatures and/or high pressures, anharmonic and nuclear quantum effects (NQEs) may significantly alter its phonon characteristics. Unfortunately, accurate inclusion of these two effects within numerical simulations typically incurs a substantial computational cost. In this work, we present a novel approach which promises to mitigate this problem. The scheme leverages the recently introduced quantum correlators approach for the extraction of anharmonic phonon frequencies from molecular dynamics data. To account for NQEs without excessive increase of the computational cost, we include nuclear quantum effects via the quantum thermal bath (QTB) method. This is the first full exploration of the use of QTB for the calculation of phonon dispersion relations. We demonstrate the noteworthy efficiency and accuracy of the scheme, and analyze its upsides and drawbacks by first considering 1-dimensional systems, and then the physically interesting case of solid neon. 
   
\end{abstract}

\maketitle

\section{Introduction}
\label{sec:introduction}

Phonons, the quantized vibrations of a crystal lattice, are fundamental to understanding a wide range of material properties. Their behavior influences diverse physical characteristics, spanning from heat transport and mechanical stability to electronic interactions and optical responses \cite{ashcroft1976solid,srivastava2022physics}. Analyzing phonon dynamics provides key insights into thermal phenomena such as heat capacity, expansion, and conductivity \cite{ziman2001electrons}, as well as structural attributes like elasticity, shear moduli, and phase stability \cite{baroni2001phonons, born1955dynamical}. Furthermore, phonons play a critical role in optical properties—governing infrared absorption and Raman scattering \cite{cardona1982light} --- while also affecting electronic properties, including bandgap modulation and electron-phonon interactions \cite{giustino2017electron}. Phonons also contribute to more exotic phenomena such as thermoelectric performance and even superconductivity \cite{tinkham2004introduction, togo2015first}.

In this article, we present a method for calculating the lattice vibrational properties of crystalline solids that enables one to account for nuclear quantum effects (NQEs) and anharmonicity beyond the level of perturbation theory. The approach consists of sequentially performing molecular dynamics (MD) simulations, extended to include (approximately) NQEs, followed by a post-processing procedure that utilizes the harvested MD trajectories to extract phonon spectra. As demonstrated by the illustrative application to solid neon, this scheme accurately captures non-trivial effects. Furthermore, its relatively low numerical cost permits one to increase the system size and temporal duration of the simulations.

Several methods can currently be employed to carry out calculations of phonon spectra in solids. The most traditional of these approaches is density functional perturbation theory (DFPT) \cite{baroni2001phonons}, which accesses phonon frequencies through interatomic force constant matrices. DFPT computes these matrices via a perturbative approach that approximates second derivatives of the total energy, hence falling under the class of harmonic approximations. A related, but arguably less sophisticated technique is the frozen phonon method (also referred to as the small displacement method or supercells approach) which approximates these energy Hessians via finite-differences \cite{weyrich1990frozen, alfe2009phon, van2002effect}. Both of these methods are only strictly valid at zero kelvin and for classical systems, or for atoms with large mass, where nuclear quantum effects such as zero-point motion can be neglected. To go beyond this, accounting for the anharmonic effects that come into play at finite temperatures and enabling a quantum mechanical treatment of the ions, a diversity of alternative approaches have been proposed. These include the self-consistent harmonic approximation \cite{hooton1955li} (SCHA) and its stochastic extension SSCHA \cite{errea2014anharmonic, bianco2017second}, self-consistent ab-initio lattice dynamics (SCAILD) \cite{souvatzis2008entropy}, and the self-consistent phonon method \cite{werthamer1970self, tadano2015self}(SCP). These methods are based on variational principles and optimize self-consistently a trial Hamiltonian with respect to phonon modes and internal degrees of freedom of the crystalline system.  All of them either rely on perturbation theory in order to include anharmonic effects, or assume the possibility of fitting the true anharmonic potential energy surface to an effective harmonic potential. If either of these assumptions are invalid, then these strategies run into difficulties.

Alternatively, there are a family of methods that utilize molecular dynamics to characterize the lattice vibrations of crystalline systems. Such MD simulations are typically ab initio (AIMD) in nature, though the use of machine-learned potentials is becoming more prevalent in the literature. The temperature-dependent effective potential (TDEP) method belongs to this category of approaches \cite{hellman2011lattice}. In TDEP, AIMD trajectories are generated, and the resulting forces are used to determine the optimal harmonic force constants. These force constants are determined by minimizing the difference between the calculated DFT forces and those predicted by a model harmonic Hamiltonian, typically using a least-squares procedure, which provides the criterion for optimality. Molecular dynamics trajectories are also at the core of the recently introduced quantum correlators (QC) approach of Ref.\citen{morresi2021probing}, which allows one to capture all orders of anharmonicity without any underlying assumptions limiting the systematic improvement of the treatment. The approach, originally developed for finite temperature calculations of generalised normal modes for classical systems, has lately been generalised to path integral molecular dynamics (PIMD)\cite{rpmd}, thus accessing essentially exact sampling of the underlying quantum Boltzmann distribution and including quantum effects in the phonon spectrum. The issue with using PIMD to incorporate nuclear quantum effects is, however, the increased numerical cost it incurs which, recent progress notwithstanding~\cite{ceriotti2012efficient, brieuc2016quantum, ceriotti2009nuclear, markland2008efficient, ceriotti2011accelerating, buxton2017accelerated, marsalek2016ab, fletcher2021fast, musil2022quantum}, may still remain problematic. To circumvent this computational expense, in this work we have adapted the QC scheme to take advantage of a recently proposed method for approximate sampling of the quantum distribution of the nuclei, namely the quantum thermal bath (QTB) \cite{qtb}. This is, to the best of our knowledge, the first time QTB has been applied to the full study of phonon dispersions (although it has been employed to assess isotope effects in LiH and LiD crystals where the authors also explored phonon densities of state \cite{dammak2012isotope}). The non-trivial exploratory calculations presented here provide the groundwork for future use of QTB to investigate phonon properties in quantum crystals.

\section{Methods}
\label{sec:Methodology-theory}

\subsection{Quantum anharmonic phonons via Kubo correlators} \label{subsec:Methods-anarmonic-phonon-theory}
In Ref.\citen{morresi2021probing} a new approach to compute anharmonic phonon spectra including quantum nuclear effects within the path integral formalism was proposed. The method determines the matrix of principal modes, $Y$, and the characteristic frequencies, $\omega_{FF}^2$, of the system by solving the following generalized eigenvalue problem (GEV):  
\begin{equation}
  \label{eq:ffpp}
  \left[ \langle \langle \boldsymbol{\mathsf{F}} \boldsymbol{\mathsf{F}}^T  \rangle \rangle \right]_{i_1 i_2} Y_{i_{2}i_{3}}=\omega_{FF,i_{3}}^{2}\left[ \langle \langle \boldsymbol{\mathsf{p}} \boldsymbol{\mathsf{p}}^T \rangle\rangle \right]_{i_1 i_2} Y_{i_{2}i_{3}}
\end{equation}
where $i_1,i_2,i_3$ are atomic indexes, $\boldsymbol{\mathsf{F}}$ are forces and $\boldsymbol{\mathsf{p}}$ momenta. The double bracket indicates that Kubo correlations are to be used. The Kubo correlation of $A$ and $B$ is defined as
\begin{equation}
    K_{AB}=\langle \langle A B \rangle \rangle = \frac{1}{\beta Q}\int_0^{\beta} d\lambda \mathrm{Tr} \left [ e^{-(\beta - \lambda) H } Ae^{-\lambda H}B\right ]
\end{equation}
where $H$ is the Hamiltonian, $\beta=1/k_B T$ ($k_B$ is the Boltzmann constant and $T$ the temperature), and $Q=\mathrm{Tr} \left [e^{-\beta H}\right ]$ is the partition function. Use of Kubo correlators ensures, in particular, that $\beta\langle \langle \boldsymbol{\mathsf{F}} \boldsymbol{\mathsf{F}}^T  \rangle \rangle $ is an estimator of the Hessian of the Hamiltonian, i.e. the dynamical matrix, and that $\langle \langle \boldsymbol{\mathsf{p}} \boldsymbol{\mathsf{p}}^T \rangle\rangle$ is diagonal and an estimator of $mk_BT$. Furthermore, the diagonal nature of the Kubo correlation for momenta ensures that the principal mode matrices are unitary in mass-weighted coordinates as expected for phonons~\cite{born1955dynamical}.

An alternative estimator for the phonon spectrum was also proposed in Ref.\citen{morresi2021probing}, based on 
\begin{equation}
\label{eq:pos-vel-gev}
  \left[ \langle\langle  \delta _{ } \boldsymbol{\mathsf{R}} \delta _{ } \boldsymbol{\mathsf{R}} ^{T}   \rangle \rangle^{-1}   \right] _{i_1 i_2} W _{i_2 i_3} = \omega_{xx,i_3} ^{2}  \left[  \langle \langle \dot{\boldsymbol{\mathsf{R}}} \dot{\boldsymbol{\mathsf{R}}} ^{T}   \rangle \rangle ^{-1}  \right] _{i_1 i_2} W _{i _{2} i _{3}} 
\end{equation}
where $\delta\boldsymbol{\mathsf{R}}$ are atomic displacements and $\dot{\boldsymbol{\mathsf{R}}}$ atomic velocities, $W$ is the matrix of principal modes for this case, and $\omega^2_{xx}$ indicates the corresponding characteristic frequencies. This generalised eigenvalue problem is equivalent to Eq.~\ref{eq:ffpp} at low temperature or small anharmonicity of the potential. It can also be shown that it provides a quantum analogue of Principle Component Analysis~\cite{martinez2006extracting}. Once again, use of Kubo correlations (and in particular the diagonality of the velocity correlation matrix) leads to a principal mode matrix that is unitary in mass-weighted coordinates. 

In implementations so far, the equations above were computed by estimating the (zero-time) correlations via PIMD trajectories. In particular, the Kubo correlations of $\delta \boldsymbol{\mathsf{R}}$ or $\boldsymbol{\mathsf{F}}$ were directly obtained from a double average over the path integral beads~\cite{morresi2021probing}, while the Kubo correlations of the momenta, proportional to $\beta$, were determined via the classical kinetic energy distribution of the beads. When computing the phonon dispersion for crystals, the simulations were performed in a periodically repeated supercell with sites $\vb{R} = \left( n_1 \vb{a}_1, n_2  \vb{a}_2, n_3 \vb{a}_3\right)$, where $\vb{a}_l$ ($l=1,2,3$) are the Bravais lattice vectors and the integers $n_l=0,..,N_l$ identify the primitive cell (for $n_l=0$) and its $N_l-1$ periodic images in the $l$-th direction. To ensure translational gauge invariance of the displacement correlation matrix, the pinning procedure detailed in Ref.~\citen{morresi2021probing} was adopted. 
In this procedure, one first defines a set of ``pinning centers'', $\vb{R}_{\tau}(t)$, which are the instantaneous positions of the symmetry-equivalent atoms within the supercell. Then the thermal average of the displacement of each of the $N$ atoms from each pinning center, $\vb{R}_{i}(t) - \vb{R}_{\tau}(t)$, is computed. Finally, correlations of the deviations of the displacements from these averages are calculated for each pinning center and averaged over the pinning centers.

 % \textcolor{red}{In summary, this procedure prescribes that for each symmetry-equivalent atom in the simulation supercell, which is referred to as a "pinning center", the displacement vector $\delta \vb{R}$ to each other atom in the supercell should be calculated with reference to that pinned atom, and the correlators averaged over all pinning positions.}\sabo{To be reformulated by introducing the pinned atom...otherwise we (I) can't say "that pinned atom"...} 

In spite of its successes, PIMD can become prohibitively expensive when the system size increases or the temperature and pressure lead to highly quantum systems and several approaches have been proposed to mitigate this problem~\cite{ceriotti2012efficient, brieuc2016quantum, ceriotti2009nuclear, markland2008efficient, ceriotti2011accelerating, buxton2017accelerated, marsalek2016ab, fletcher2021fast, musil2022quantum}. In the following, we explore an alternative and less computationally demanding means of incorporating NQEs into simulations: the quantum thermal bath (QTB)~\cite{qtb, ceriotti2009nuclear}.
% In addition to this pinning strategy, we found that it was particularly important to remove the effect of global translations 

\subsection{QTB trajectories and Kubo correlators}
 QTB mimics nuclear quantum thermal effects by combining classical dynamics of the system with quantum thermalization in a generalized Langevin evolution:

\begin{align}
\label{qtb_eom}
\begin{split}
        m_I \dv[2]{\vb{R}_I}{t} = - \nabla_{\vb{R}_I} V(\{\vb{R}_I\}) - m_I \gamma \dv{\vb{R}_I}{t} + \boldsymbol{\xi}_I(t)
        \end{split}
\end{align}
where $\gamma$ is the friction coefficient, and $\boldsymbol{\xi}_I = [\xi_I^{x}\  \xi_I^y \ \xi_I^z]$ is a 3-dimensional random force vector associated with ion $I$, with $I$ going from 1 to the number of ions. $V(\{\vb{R}_I\})$ is the interaction potential, possibly obtained ab initio or via machine learning. The random force is endowed with the coloured noise spectrum 
 \begin{equation}
 \label{qtb_fdt}
        C _{\xi^{\alpha}_J \xi^{\beta}_K} (\omega) = 2m_K \gamma \theta(\omega, T) \delta _{JK} \delta ^{\alpha \beta} \qquad \alpha, \beta \in \{x,y,z \}
    \end{equation}
where $C _{\xi^{\alpha}_J \xi^{\beta}_K}$ is the autocorrelation function of the $\alpha$'th component of the force on ion $J$ and the $\beta$'th component of the force on ion $K$, $m_K$ is the mass of the K'th ion, and 
\begin{equation}
    \theta(\omega, T) = \frac{\hbar \omega}{2\tanh(\frac{\omega}{2 k_B T})}
\end{equation}.

The QTB method can be justified within the framework of Feynman-Vernon influence functional theory~\cite{kleinert1995quantum}. In this context, the QTB evolution equation arises in a system-harmonic bath model, expanding the influence functional to first order in the fluctuation variables entering into its forward-backward path integral representation. The required hypothesis is that the system is weakly coupled to the bath of harmonic oscillators via an Ohmic dissipative coupling which is Markovian in nature (local-in-time)~\cite{schmid1982quasiclassical}.
 The QTB evolution, which is exact for a quantum harmonic oscillator, is only approximate for more general interactions. However, it has proved to be surprisingly effective in several applications and is currently being explored and exploited in different contexts as a computationally cheaper alternative to PIMD \cite{bronstein2014quantum, mauger2021nuclear,qtb, dammak2012isotope, hernandez2015applicability, bronstein2014quantum, bronstein2017thermal,bronstein2016quantum}.
One well-known drawback of the quantum thermal bath, which is common to a number of semiclassical approaches, is zero-point energy leakage (ZPEL) \cite{brieuc2016zero}. This is the phenomenon wherein energy stored in higher frequency modes, thermalized according to the quantum thermal distribution generated by QTB, flows unphysically into lower frequency modes when the system is propagated via classical dynamics in the presence of anharmonicity and couplings. This problem can be largely mitigated by adopting a recently proposed scheme known as the adaptive quantum thermal bath (adQTB) \cite{mangaud2019fluctuation}. A summary of this method, and of important aspects of the algorithmic implementation of QTB or adQTB runs is presented in Appendix~\ref{appendix:QTBadQTB}. In the following, we focus instead on a specific technical point that must be addressed to employ QTB trajectories to obtain quantum anharmonic phonons.
 
QTB trajectories were originally introduced to perform an approximate sampling of the quantum thermal distribution and compute static quantities. Since the trajectories provide simultaneous values for momenta and coordinates of the system, the QTB distribution is often associated with the Wigner representation of the quantum probability~\cite{basire2013computing}. More recently, in analogy with calculations based on PIMD, these trajectories have also been used, with surprising accuracy, to obtain time-dependent observables, such as vibrational spectra. Averages, and in particular, correlation functions (time-dependent or at zero time) computed via QTB, however, are typically associated with standard and not Kubo quantum averages. Standard time-correlation functions are defined as
\begin{equation}
    C_{AB}(t) = \langle AB(t)\rangle=\frac{1}{Q}\mathrm{Tr} \left [ e^{-\beta H}A B(t) \right ]
\end{equation}
To connect with the Kubo correlators required in the phonon calculations, we can exploit the relationship between the Fourier transform of the standard
and Kubo time-correlation functions. This relationship is given by 
\begin{equation}
\label{eq:KuboVsStd}
    \tilde{K}_{AB}(\omega)  = \frac{2}{\beta \omega}\mathrm{th}\left(\frac{\beta \omega}{2}\right)\mathrm{Re}\left[\tilde{C}_{AB}(\omega)\right] 
\end{equation}
Using QTB we can compute an estimator of the full time-dependent correlation function $C_{AB}(t)$ and then obtain its Fourier transform. The Kubo-transformed zero-time correlation is then computed upon integration over the frequency domain of the right hand side of the equation above. Note that, although we ultimately make use of the zero-time Kubo-transformed TCF, our calculation requires the evaluation of the full TCF (not just at $t=0$). This results in a computational overhead (both in terms of memory and computing time) compared to post-processing in phonon calculations via PIMD. However, by leveraging a batching strategy to reduce the memory requirement and parallelizing the computation of the different components of the correlation matrices, it is possible to mitigate this overhead to a large extent.% The downside being, of course, the requirement of supplementary computational resources. %To circumvent both the need for additional computational resources and the increased memory requirements, we have also explored an alternative strategy which involves approximating the zero-time Kubo-transformed TCF by the zero-time standard TCF. The justification for such an approximation is now outlined.

\section{Tests on one-dimensional models}
\label{sec:toy_models}
A preliminary assessment of using QTB in tandem with the QC methodology was carried out by considering three one-dimensional models. These models are a meaningful subset of those introduced in Ref.~\cite{morresi2021probing}: 
\begin{enumerate}
  \item A harmonic oscillator 
\begin{equation}
    \label{eq:harm_pot}
  V(x)=\frac{k}{2}x^2 \qquad k=0.183736 \text{ a.u.} 
\end{equation}

    \item A ``mildly anharmonic'' Morse potential, given by \eqnref{eq:morse_pot}, with $a_m=0.2 \text{ bohr}^{-1}$.
    \item A ``strongly anharmonic'' Morse potential, given by \eqnref{eq:morse_pot}, with $a_m=0.8 \text{ bohr}^{-1}$.
\begin{align}
  \label{eq:morse_pot}
  V(x)=\frac{k}{2a_m^2}(1-e^{-a_m x})^2 \qquad &k=0.183736 \text{ a.u.},\\
  \nonumber
  &a_m=0.2,0.8 \text{ bohr}^{-1}
\end{align}

\end{enumerate}

The harmonic case was considered to verify the proposed approach in a case in which QTB provides exact sampling. The two values of the parameter $a_m$ in the Morse potential, on the other hand, allowed us to investigate the ability of QTB to reproduce accurate frequencies when different levels of anharmonicity are present in the system. To quantify the degree of the anharmonic effects due to the choice of $a_m$, we adopted a recently introduced measure for anharmonicity in material systems, developed by Knoop \etal \cite{knoop2020anharmonicity}. This metric is based on the ratio of the root mean square of the anharmonic contribution to the force acting in the system to the total force: $\sigma^A  = \frac{\sigma[\mathbf{F}^A]}{\sigma[\mathbf{F}]} = \sqrt{ \frac{ \langle ( \mathbf{F} - \mathbf{F}^{(2)} )^2 \rangle }{ \langle \mathbf{F}^2 \rangle } }$, where $\mathbf{F}^A = \mathbf{F} - \mathbf{F}^{(2)}$ is the difference between true and harmonic forces. We find that the choices of $a_m=0.2$ and $a_m=0.8$ yield values of $\sigma^A\approx 0.084$ and $\sigma^A\approx 0.370864$, with the latter indicating that the level of anharmonicity present for $a_m=0.8$ is indeed significant according to the general observations of Knoop \etal, who found values of $\sigma^A \gtrapprox 0.2$ already indicate the onset of noticeable anharmonic effects.

The frequencies for these models are obtained by solving the one-dimensional versions of \eqnref{eq:ffpp} (FF-pp results in the following) and \eqnref{eq:pos-vel-gev} (xx-vv results in the following) for the square root of the ``eigenvalue'' $\omega^2$. For all systems, we performed classical, and QTB simulations. Reference PIMD results were taken from Ref.~\citen{morresi2021probing}. For the classical case, which can be seen as the single-bead version of the path integral procedure, the correlators were computed via averages along classical Langevin trajectories. The timestep employed in these simulations was set to dt=0.1 fs. The friction coefficient was set to $\gamma=10^{-3}$ a.u. in the classical case and $\gamma=10^{-4}$ a.u. for QTB. The cutoff frequency was set to $\omega_{cut}=0.03$ a.u. for the QTB simulations. These parameters are appropriate choices for the characteristic frequencies ($\omega\sim 2100$ \cminv{}) of the systems considered. 
In all tests, the mass of the simulated particles was set equal to that of the hydrogen atom, m=1837.154 a.u., and the temperature was set to 20K. 

In \tabref{tab:harmonic_frequencies}, we present the results for the harmonic system. In addition to the results for the Kubo-transformed QTB correlations, we show (in row three of the table) also the frequencies obtained via direct averages along the QTB trajectories. As discussed previously, these averages correspond to the standard quantum correlation function. They are reported here and in the following to assess the importance of accounting for the reweighting induced by \eqnref{eq:KuboVsStd}. As expected for this system, PIMD, QTB, and classical results are essentially indistinguishable for both FF-pp and xx-vv correlations. Moreover, as expected for a harmonic potential, the QTB frequencies obtained via use of either standard or Kubo-transformed TCFs are in virtually perfect agreement. 

%\begin{equation}
%  \label{eq:ffpp2}
%  \left\langle F^2\right\rangle Y=\omega^2 \left\langle p^2\right\rangle Y
%\end{equation}
%\begin{equation}
%  \label{eq:xxvv2}
%    \left\langle (\delta x)^2\right\rangle^{-1} W=\omega^2 \left\langle \dot{x}^2\right\rangle^{-1} W
%\end{equation}. 

%\subsection{Results and discussion}
%\label{sec:toy_model_results}
\begin{table}[H]
\centering

\begin{tabular}{|c|c|c|}
\hline
Method & xx-vv & FF-pp \\
\hline
PIMD & 2194.74${}^a$ & 2194.74${}^a$ \\
QTB (Kubo-TCF) & 2194.08 & 2195.66 \\
QTB (Std-TCF) & 2195.00 & 2194.74 \\
Classical & 2194.44 & 2195.66 \\
\hline
\end{tabular}
\caption[Table of computed frequencies for the harmonic potential given in \eqnref{eq:harm_pot}.]{Table of computed frequencies for the harmonic potential given in \eqnref{eq:harm_pot}. ${}^{a}$ Value taken from Ref.\citen{morresi2021probing}.}
\label{tab:harmonic_frequencies}
\end{table}

\begin{figure*}[t]
  \centering
    \centering
    \includegraphics[scale=0.5]{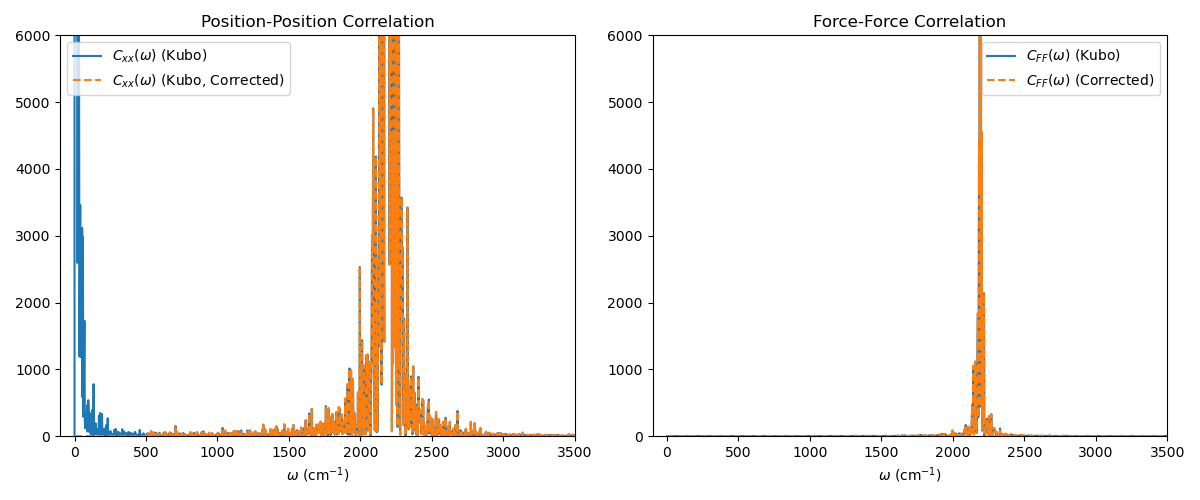}
    \caption{(Left) Plot of the Kubo-transformed displacement-displacement correlation function in the frequency domain for the mildly anharmonic Morse potential. The blue line depicts the usual centered, Kubo-transformed correlation function in the frequency domain, while the orange line shows the Kubo-transformed correlation function with application of the correction procedure described in the main text. (Right) Analogous plot for the Kubo-transformed force-force correlation function.}
    \label{fig:dos_both}

    % \caption{Plot of the Kubo-transformed force-force correlation function for the mildly anharmonic Morse potential. The blue line depicts the usual centered, Kubo-transformed correlation function in the frequency domain, while the orange line shows the same correlation function, but with application of correction procedure described in the main text.}
    % \label{fig:ff_kubo}
\end{figure*}

\tabref{tab:morse_low_anharm} reports the frequencies computed for the Morse potential in the case of mild anharmonicity.
Let us begin by considering the FF-pp correlators, ignoring - for the time being - the row noted as Kubo-TCF, Corrected. In this case too, there is very good agreement between the PIMD and Kubo-transformed QTB results, while the classical frequency is slightly blue shifted. The non-Kubo-transformed result is close but not coincident with the fully quantum reference. The results for the xx-vv correlator are more interesting in that there is a relevant discrepancy between the PIMD and Kubo-transformed QTB frequencies. The cause of this discrepancy is illustrated %a non-zero average of the position of the system that manifests itself as a peak at $\omega=0$ in the corresponding power spectrum. The peak is further exacerbated by the Kubo kernel,  which is itself sharply peaked towards zero frequency, and which instead weights to a much lesser extent signals at higher frequencies. This artifact is trivially eliminated by subtraction of the average in the displacement correlation: $\langle (\delta x(t))^2 \rangle = \langle (x(t)-\langle x \rangle)^2\rangle$.  In practice, the effects of the translations are removed by setting to zero the $\omega=0$ component of the Fourier transform of the correlation, along with an envelope of frequencies centred around zero but extending to a finite range $\Delta \omega$. This broadening accounts for the effects of the Langevin thermostat. For the one-dimensional models considered here (including the strongly anharmonic Morse potential), the choice of the frequency range to be encompassed by this envelope is simple given the clear separation of the physically relevant modes near the natural characteristic frequency of the potential and these unphysical modes localized around $\omega = 0$. This discussion is further illustrated 
in \figref{fig:dos_both}. The displacement-displacement density of states (DOS) plotted on the left-hand-side of this figure, shows that even with centering of the correlation function (i.e., subtracting $\langle x\rangle$) as prescribed by Eq.~\ref{eq:pos-vel-gev}, there is a residual signal around zero frequency which arises from thermal broadening of the signal due to the thermostat. Given the shape of the Kubo kernel in frequency space, which tends to unity as $\omega\to0$, this anomalous signal remains significant in the resulting Kubo-transformed correlation function. On the other hand, the physically meaningful signal at around $2194$ \cminv is diminished by the rapidly decaying behaviour of the kernel towards large frequencies. This produces the large discrepancy observed for the Kubo-transformed correlation function. Contrariwise, the fact that the standard correlation function does not assign such skewed weights to different frequencies explains why our results using standard TCFs appear much less susceptible to the presence of spurious peaks at low frequency.  To mitigate this problem, we set to zero the $\omega = 0$ component of the Fourier transform of the xx correlation, along with
an envelope of frequencies centred around zero and extending to a finite range $\Delta \omega$ to counteract the effects of
the Langevin thermostat. For the one-dimensional Morse models considered here, the choice of the frequency range to be encompassed by this envelope is simple given the clear
separation of the physically relevant modes near the natural characteristic frequency of the potential and these
unphysical modes localized around $\omega = 0$. Results after this correction are reported in the fourth row of Table~\ref{tab:morse_low_anharm}, and show marked improvement. Figure~\ref{fig:dos_both} (right panel) clarifies also why the FF-pp results do not show the same problem. Because the velocities are time derivatives of the positions and forces are the time derivatives of the momenta, the Fourier transform of the velocity correlation function is equal to that of the position correlation function multiplied by $\omega^2$, and similarly for the Fourier transform of the force correlation function with respect to the momentum correlation function. As a result, there is a factor $\omega^4$ between $C_{xx}(\omega)$ and $C_{FF}(\omega)$, be they the standard or Kubo-transformed correlation functions \cite{kubo2012statistical}. This factor decays quickly to zero as $\omega \rightarrow 0$, such that the spurious low frequency peak is suppressed  for $C_{FF}(\omega)$ as shown in \figref{fig:dos_both}. 

% \begin{figure*}[t]
%   \centering
%   \begin{minipage}{0.48\textwidth}
%     \centering
%     \includegraphics[width=\textwidth]{figures/posposm_kubo.png}
%     \caption{Plot of the Kubo-transformed displacement-displacement correlation function for the mildly anharmonic Morse potential. The blue line depicts the usual centered, Kubo-transformed correlation function in the frequency domain, while the orange line shows the Kubo-transformed correlation function with application of the described correction procedure. \tb{Single caption or stick with two?}}
%     \label{fig:posposm_kubo}
%   \end{minipage}
%   \hfill
%   \begin{minipage}{0.48\textwidth}
%     \centering
%     \includegraphics[width=\textwidth]{figures/ff_kubo.png}
%     \caption{Plot of the Kubo-transformed force-force correlation function for the mildly anharmonic Morse potential. The blue line depicts the usual centered, Kubo-transformed correlation function in the frequency domain, while the orange line shows the same correlation function, but with application of correction procedure described in the main text.}
%     \label{fig:ff_kubo}
%   \end{minipage}
% \end{figure*}

\begin{table}[H]
\centering

\begin{tabular}{|c|c|c|}
\hline
Method & xx-vv & FF-pp \\
\hline
PIMD & 2190.86${}^a$ & 2193.55${}^a$ \\
\hline
QTB (Kubo-TCF)& 2049.80 &  2194.63 \\
%QTB (Kubo-TCF)& 1911.06 & 2193.56 \\
\hline
QTB (Kubo-TCF, Corrected) & 2186.53 & 2193.56 \\
\hline
QTB (Std-TCF) & 2194.83 & 2187.64 \\
\hline
Classical & 2194.54 & 2194.74 \\
\hline
\end{tabular}
\caption[Table of computed frequencies (given in units of \cminv{}) for the morse potential given in \ref{eq:morse_pot} for $a_m=0.2 \text{ bohr}^{-1}$ (mild anharmonicity).]{Table of computed frequencies (given in units of \cminv{}) for the Morse potential given in \eqnref{eq:morse_pot} for $a_m=0.2 \text{ bohr}^{-1}$ (mild anharmonicity). ${}^a$ Value taken from Ref.\citen{morresi2021probing}. }
\label{tab:morse_low_anharm}
\end{table}

We now move to the results in~\tabref{tab:morse_high_anharm} where the frequencies computed for the Morse potential in the case of strong anharmonicity are reported, starting again with the FF-pp correlators. As expected for this more challenging case for QTB, the agreement between the QTB-Kubo and PIMD frequency degrades somewhat. However, the method correctly captures the red shift of the frequency compared to the classical calculation that remains essentially identical to the result obtained for the mildly anharmonic case. We note that the non-Kubo-transformed QTB appears to be closer to the reference result, pointing to some residual instabilities of the numerical Kubo transform. This is highlighted in the xx-vv results that show a rather dramatic error for the brute-force QTB-Kubo. This is clearly mitigated by our correction procedure that strongly reduces the difference with the reference result.

\begin{table}[H]
\centering
\begin{tabular}{|c|c|c|}
\hline
Method & xx-vv & FF-pp \\
\hline
PIMD & 2132.70${}^a$ & 2175.55${}^a$ \\
\hline
QTB (Kubo-TCF) &  1185.26 &  2149.34\\

%QTB (Kubo-TCF) & 962.50 & 2160.53 \\
\hline
QTB (Kubo-TCF, Corrected) & 2092.16 & 2160.53 \\
\hline
QTB (std-TCF) &  2096.25 & 2169.40 \\
%QTB (std-TCF) & 2061.52 & 2169.84 \\
\hline
Classical & 2195.88 & 2195.88 \\
\hline
\end{tabular}
\caption[Table of computed frequencies (given in units of \cminv{}) for the morse potential given in \ref{eq:morse_pot} for $a_m=0.8 \text{ bohr}^{-1}$ (significant anharmonicity).]{Table of computed frequencies (given in units of \cminv{}) for the Morse potential given in \ref{eq:morse_pot} for $a_m=0.8 \text{ bohr}^{-1}$ (significant anharmonicity).${}^a$ Value taken from Ref.\citen{morresi2021probing}.}
\label{tab:morse_high_anharm}
\end{table}

Some general remarks can be made on the basis of these tests. When the FF-pp correlator is adopted, QTB-Kubo is in good agreement with the PIMD results, with a maximum discrepancy of about 25 cm$^{-1}$ when significant anharmonicity is introduced in the model. Even in this situation, our approach correctly captures the behaviour of the quantum frequencies relative to the classical results, in particular with respect to the red shift resulting from the quantum effects. In contrast, the classical calculations are essentially insensitive to the change of model, most likely due to missing zero point energy effects. In fact, at 20K the classical particle remains near the bottom of the potential well, which can be well approximated as being harmonic (note that the harmonic approximation of the Morse potential has the same curvature as our purely harmonic system). On the other hand, QTB captures quite well the zero point motion that allows the particle to explore a larger portion of the potential energy surface. The results reported for the xx-vv correlator highlight a potential numerical instability of our approach when frequencies close to $\omega = 0$ are present. Small errors in the Fourier transform of the QTB correlation functions around those frequencies can be strongly amplified by the relationship in Eq. \ref{eq:KuboVsStd}, leading to serious errors as demonstrated by the results in \tabref{tab:morse_low_anharm} and \tabref{tab:morse_high_anharm}. The correction procedure that we have outlined is effective, but it entails zeroing components of the Fourier transform of the standard QTB correlation over a range of frequencies. The correct choice for this frequency envelope is problematic, for example, when acoustic phonon modes that give rise to physically relevant frequencies in the region near $\omega=0$ are present. In the next section, we propose a scheme to circumvent, to some extent, this difficulty. Our test calculations also show that the non-Kubo-transformed QTB correlators lead to the computed frequencies exhibiting qualitatively correct trends.

\section{Solid neon}
\label{sec:neon}
We now present results obtained for the vibrational modes of Lennard-Jones neon in its solid phase. As the lightest noble solid after helium, neon presents an excellent test case due to a combination of its properties. While helium's crystalline phase is difficult to study because its Debye temperature greatly exceeds its very low melting point, neon remains solid up to $\sim$25 K with a Debye temperature of $\sim$75 K\cite{tari2003specific}. Moreover, solid neon exhibits significant anharmonicity and pronounced NQEs at low temperatures \cite{ramirez2008path,neumann2000path,endoh1975lattice}. These attributes make it an ideal benchmark for assessing the effectiveness of our approach on a realistic model of a condensed phase system.

The system was modelled as a supercell containing 64 neon atoms arranged in a face-centered cubic (FCC) structure. The lattice constant was chosen to be 4.48 \AA, and the temperature was maintained at 15.185 K. The mass of each neon atom was set to 20.1797 Da, reflecting the weighted average of the naturally occurring isotopic distribution of neon. Interactions were described via the Lennard-Jones potential:
\begin{equation}
    V(r) = 4\epsilon \left[\left(\frac{\sigma}{r}\right)^{12} - \left(\frac{\sigma}{r}\right)^{6}\right]
\end{equation}
with $\epsilon = 3.16$ meV and $\sigma = 2.79$ \AA. Using an empirical force field allowed us to perform several PIMD simulations at relatively low numerical cost, providing benchmarks for the QTB-based calculations. Furthermore, the Lennard-Jones potential provides a reliable approximation of neon's interatomic interactions, enabling also comparisons with experiments. 

Simulations were performed using a timestep of 1 fs, and trajectories were generated for a total duration of 60 ps. To allow for system equilibration, the first 2 ps of each trajectory were excluded from the analysis. This simulation length was sufficient to ensure that the phonon frequencies converged to within 0.1 \cminv{}.
Classical MD simulations were conducted using an in-house code implementing the BAOAB\cite{leimkuhler2013,leimkuhler2015molecular} algorithm for Langevin dynamics. QTB simulations utilised the same software, leveraging an additional module specifically designed to perform QTB dynamics. Neon exhibits a Debye frequency of $\omega_D \approx 1.56 \times 10^{12}$ Hz, or $3.77 \times 10^{-5}$ a.u., which represents the highest vibrational frequency in the system. A cutoff frequency of $\omega_{\text{cut}} = 10^{-3}$ a.u. was then applied in the QTB simulations (see Appendix for a discussion on this parameter). To validate this choice, the average kinetic energy from QTB simulations was compared to results from converged PIMD simulations. A friction parameter of $\gamma = 2 \times 10^{-5}$ a.u. was selected to effectively suppress ZPEL while minimizing broadening of the QTB response function. Absence of ZPE was verified by monitoring the fluctuation-dissipation relation for the system\cite{mangaud2019fluctuation}. The noise was updated every 1000 steps, equivalent to 1 ps. Path-integral molecular dynamics (PIMD) simulations were conducted using the i-PI python interface \cite{litman2024pi}. The PIMD runs employed 32 beads, which provided sufficient convergence. Temperature control was achieved using i-PI’s local PILE thermostat \cite{ceriotti2010efficient}, with the thermostat time constant set to 100 fs.

\subsection{Phonon dispersions}
\label{sec:neon_dispersion_results}
\begin{figure}[H]
    % include first image
        \centering
  \includegraphics[width=0.5\textwidth]{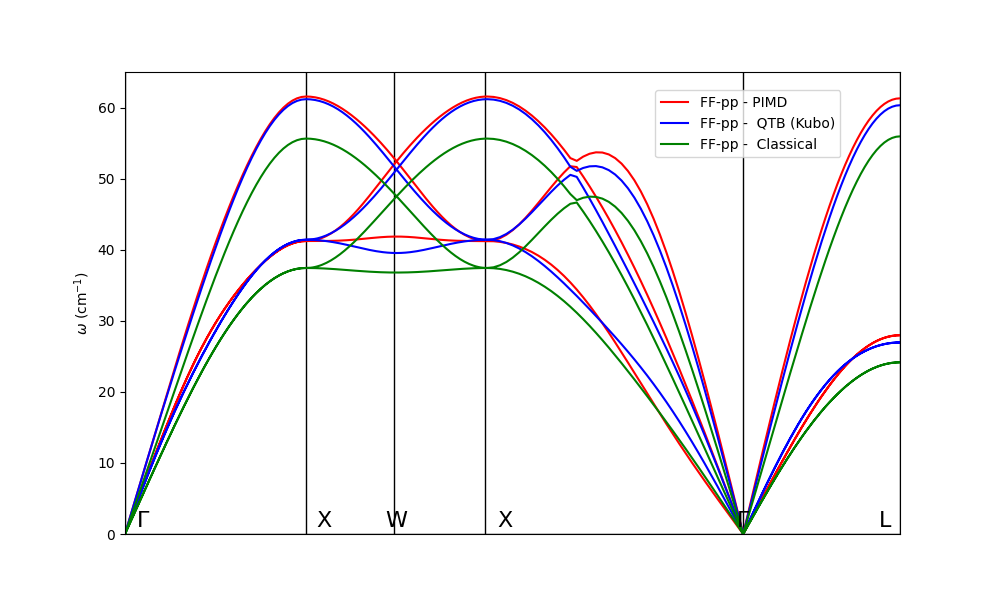}
  % include second image
   \caption[Phonon dispersions for Lennard-Jones neon at 15.185K using PIMD (red curves), QTB (blue curves), and classical MD (green curves).]{Phonon dispersions for Lennard-Jones neon at 15.185K using PIMD (red curves), QTB (blue curves), and classical MD (green curves), computed using the FF-pp GEV.}
  \label{fig:neon_lj_15k_ff}
\end{figure}

We begin by reporting results for the FF-pp GEV. Figure~\ref{fig:neon_lj_15k_ff} shows phonon dispersions for classical (green curves), PIMD (red curves) and QTB-Kubo (blue curves) calculations. There is a significant shift in the frequencies obtained from the classical MD simulations when compared to the PIMD and QTB results, evidencing the importance of nuclear quantum effects in the system. Moreover, even at this low temperature, the classical results deviate significantly from the harmonic reference (see Ref.\citen{efremkin2022study}), indicating the presence of a considerable degree of anharmonicity that the QC approach is clearly capable of capturing. Both our classical and PIMD results are in good agreement with the theoretical frequencies reported for the same system by Efremkin \etal{} in Ref.\citen{efremkin2022study}, and the experimental results of Endoh \etal{} in Ref.\citen{endoh1975lattice}. The QTB-Kubo dispersions are in very good agreement with the quantum reference, except for small discrepancies around the W point and the maximum between the X and $\Gamma$ points. These are, however, much smaller than the differences between classical and PIMD curves, indicating that QTB provides a good approximation for this system.

% \begin{figure}[H]
%     % include first image
%       \includegraphics[width=1.0\linewidth]{figures/phonon_dispersion_comp_qtb_kubo_std_class_pimd_ff.png}  
%   \includegraphics[width=1.0\linewidth]{figures/phonon_dispersion_comp_qtb_kubo_std_class_pimd_xx.png}  
%   % include second image
%    \caption[Phonon dispersions for Lennard-Jones neon at 15.185K using PIMD (red curves), QTB with Kubo-transformed time-correlation functions (blue curves), QTB with standard time-correlation functions (orange curves), and classical MD (green curves).]{Phonon dispersions for Lennard-Jones neon at 15.185K using PIMD (red curves), QTB with Kubo-transformed time-correlation functions (blue curves), QTB with standard time-correlation functions (orange curves), and classical MD (green curves). The upper panel shows the results obtained using the FF-pp GEV, while the lower panel shows the results obtained using the xx-vv GEV.}
%   \label{fig:neon_lj_15k}
% \end{figure}

% \begin{figure}[H]
%     % include first image
%   \includegraphics{figures/phonon_dispersion_comp_Ne_15k_xx.png}
%   % include second image
%    \caption[Phonon dispersions for Lennard-Jones neon at 15.185K using PIMD (red curve), QTB (blue curve), and classical MD (green curve).]{Phonon dispersions for Lennard-Jones neon at 15.185K using PIMD (red curve), QTB (blue curve), and classical MD (green curve), computed using the xx-vv GEV.}
%   \label{fig:neon_lj_15k}
% \end{figure}

\begin{figure}[H]
    \centering
  \includegraphics[width=0.5\textwidth]{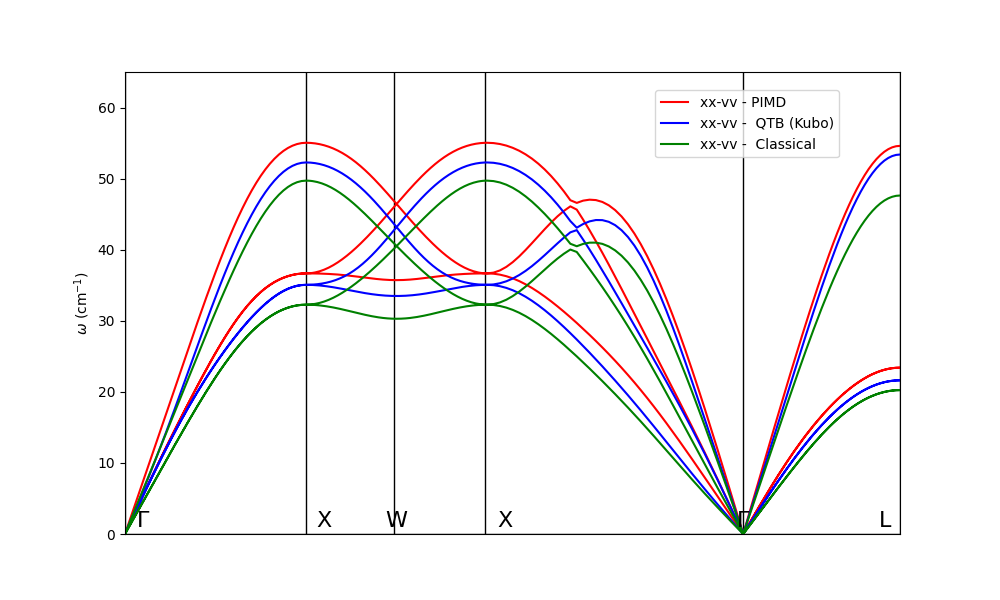}
  % include second image
   \caption[Phonon dispersions for Lennard-Jones neon at 15.185K using PIMD (red curves), QTB (blue curves), and classical MD (green curves).]{Phonon dispersions for Lennard-Jones neon at 15.185K using PIMD (red curves), QTB (blue curves), and classical MD (green curves), computed using the xx-vv GEV.}
  \label{fig:neon_lj_15k_xx}
\end{figure}

\begin{figure}[H]
    % include first image
        \centering
  \includegraphics[width=0.5\textwidth]{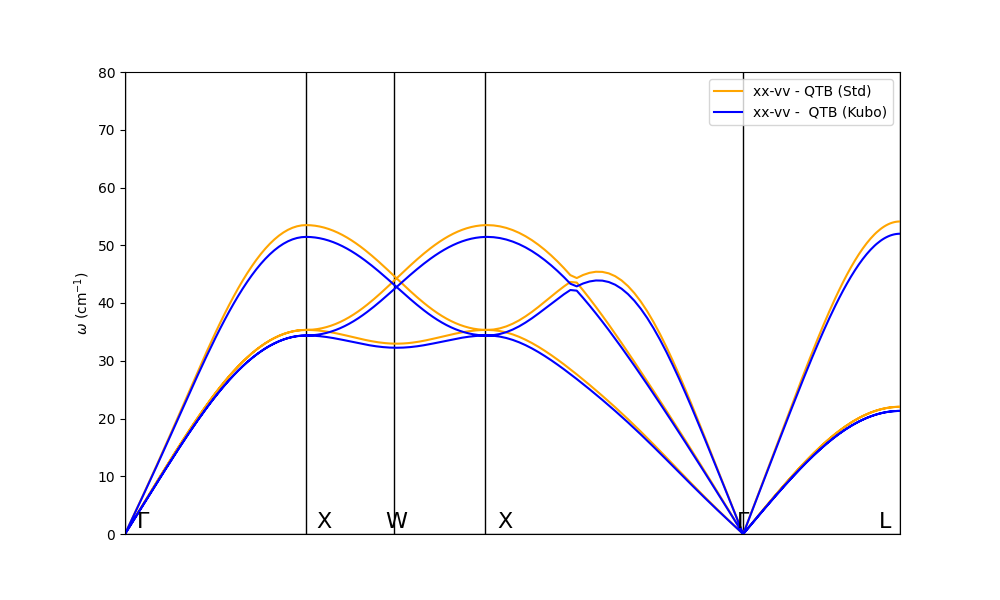}
  % include second image
\caption[Comparison of phonon dispersions for Lennard-Jones neon at 15.185K using QTB with standard time correlation functions and using Kubo-transformed time correlation functions.]{Comparison of phonon dispersions for Lennard-Jones neon at 15.185K using QTB simulations with Kubo-transformed (blue curves) and with standard (orange curves) time correlation functions. The results were obtained using the xx-vv GEV.}
  \label{fig:neon_lj_15k_xx_kubo_vs_std}
\end{figure}

\begin{figure}[H]
    % include first image
        \centering
  \includegraphics[width=0.5\textwidth]{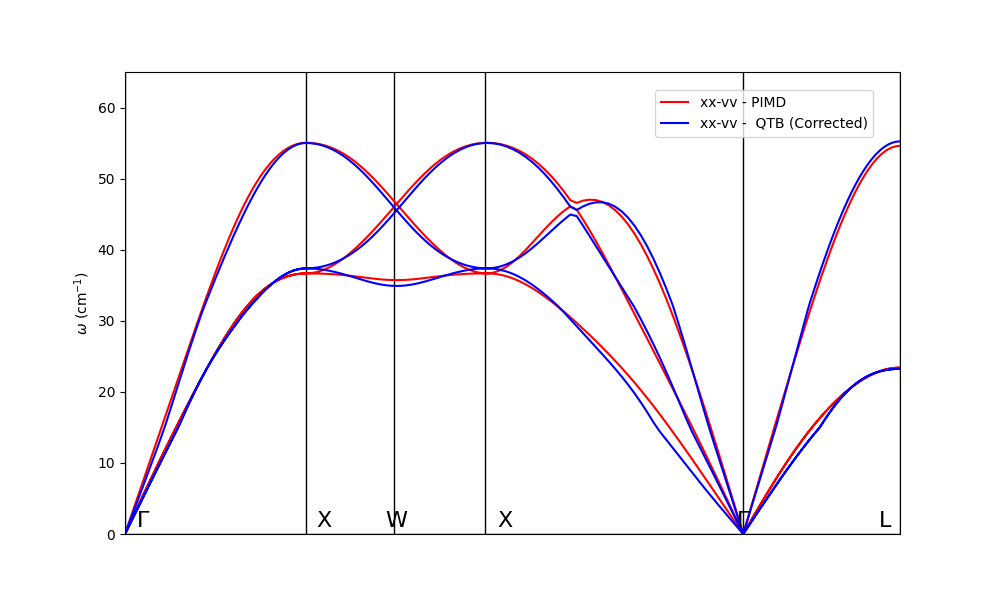}
  % include second image
\caption[Comparison of phonon dispersions for Lennard-Jones neon at 15.185K using PIMD (red curves) and QTB together with the interpolation strategy described in the main text (blue curves). ]{Comparison of phonon dispersions for Lennard-Jones neon at 15.185K using PIMD (red curves) and QTB together with the interpolation strategy described in the main text (blue curves).}
  \label{fig:neon_lj_15k_xx_corrected}
\end{figure}

In \figref{fig:neon_lj_15k_xx}, we present the phonon dispersions calculated via the xx-vv GEV (same color code as Figure~\ref{fig:neon_lj_15k_ff}). Also in this case, classical and PIMD results show considerable differences, confirming the importance of quantum nuclear effects for the system in these conditions. The agreement between QTB-Kubo and PIMD is less accurate than in the case of the FF-pp GEV dispersions. QTB underestimates frequencies along the entirety of the high symmetry path considered. Nonetheless, the qualitative agreement is still very good, and the consistent blue shift of the quantum frequencies relative to the classical frequencies is well captured. These difficulties are reminiscent of those encountered for the model systems, even though their origin from spurious peaks near $\omega=0$ in the displacement-displacement power spectral density is not as clear-cut. Indeed, the presence of physically meaningful signal around $\omega=0$ arising from the acoustic bands complicates the analysis presented in the previous section and the straightforward application of the correction procedure proposed for the Morse potentials. To improve the situation, we start by examining the discrepancy between the dispersions calculated via the QTB Kubo-transformed and standard time correlation functions. Results are shown in Figure~\ref{fig:neon_lj_15k_xx_kubo_vs_std}. The plot shows that the two results are very close, with the standard-QTB dispersions (orange curves in the plot) slightly blue shifted compared to the Kubo-transformed results at higher frequencies. The agreement between the two is however excellent (as expected from Equation~\ref{eq:KuboVsStd}) at lower frequencies, in this case up to about 25~\cminv{}. We leverage this property to propose a general empirical correction procedure for the xx-vv GEV dispersions. In this procedure, low frequency instabilities are removed by filtering out frequency components of the Kubo-transformed xx and vv TCFs before solving the GEV. This, however, introduces a systematic error because physically relevant acoustic frequencies are affected. To correct for this, we substitute the low frequency part of the Kubo-transformed dispersion with the one computed from the standard correlation function. Smooth matching of the two curves is obtained via an interpolation in the cross-over frequency region where the standard and Kubo dispersions start to differ. For neon, the filtering threshold in the calculation of the Kubo TCFs was set to $\sim$ 3~\cminv{} (this choice removes the spurious signal and empirically minimizes the impact of neglecting physically meaningful acoustic frequencies). The interpolation between standard and Kubo dispersions, on the other hand, was made in the region around 25~\cminv{} (see Figure~\ref{fig:neon_lj_15k_xx_kubo_vs_std}). The corrected-QTB results are presented in Figure~\ref{fig:neon_lj_15k_xx_corrected} and show a clear improvement in the agreement with the PIMD results. 
%Overall, the results presented in this section indicate that QTB is a promising tool for phonon calculations, capable of effectively capturing the influence of NQEs and anharmonicity on the vibrational properties of quantum crystals, of which neon is a prototypical example. 

\subsection{Comparison of computational cost of the employed methods}
\label{sec:performance_neon}
We now report the computational cost associated with the different methods used in this work. Table. \ref{tab:performance_comparison_neon} provides a breakdown of the resource requirements and run times of the different approaches. In all cases, performance is reported for Intel(R) Xeon(R) Platinum 8360Y processors running at 2.4 GHz, 512GB.
The third column gives the wall time required to generate 1 MD step using the method indicated in the first column. The ``total wall time'' column estimates the compute time required to generate a trajectory of length 60 ps, sufficient for converging phonon spectra. QTB requires essentially the same amount of computational effort as classical MD, with both performing a single MD step in $\sim$0.05 s using a single CPU. On the other hand, PIMD requires approximately 5 times the amount of wall time to achieve a single step even when 32 CPUs are employed, with individual processors assigned a given replica of the system. It should be remarked that for the present investigation we have made use of a completely standard implementation of PIMD, without exploring the potential to reduce the computational cost of these simulations by ring polymer contraction (RPC) \cite{markland2008efficient} or higher-order propagator splitting techniques \cite{takahashi1984monte,suzuki1995hybrid,chin1997symplectic,perez2011improving}, or by combination with generalized Langevin equation thermostatting (either by means of a PIGLET approach \cite{ceriotti2012efficient} or by coupling QTB with PIMD through a PIQTB-like scheme \cite{brieuc2016quantum}). This would be an interesting avenue for future study as these approaches facilitate a more systematic treatment of NQEs compared to QTB, given the ability to converge to exact sampling of the quantum Boltzmann distribution by increasing the number of replicas employed. It should also be mentioned that the post-processing overhead needed to obtain the Kubo zero-time correlations in QTB strongly reduces the difference in the overall production times for the dispersions presented in this work. However, for AIMD based calculations, where the numerical effort is concentrated in the MD rather than the post-processing, the computational resources and cost of PIMD-based simulations should remain higher than when one simply employs QTB. Furthermore, the current approach has the benefit of requiring a minimal amount of up-front parameter tuning in comparison with more advanced PIMD methods. As it stands, our computational cost assessment, together with the satisfactory results shown in the previous sections, indicate the potential for QTB to facilitate larger-scale studies of lattice vibrational properties of quantum crystals. We perform one such investigation in the following section, turning our attention to the temperature dependence of neon's phonon frequencies. 

\begin{table}[H]
    \centering
    \begin{tabular}{c|c|c|c}
         Method&Resources&\multicolumn{1}{|p{2cm}|}{\centering Wall Time\\/MD step (s)}& Total wall time (hr)\\
         \hline
         Classical&1 CPU&0.047230 &0.787 \\
         QTB&1 CPU&0.046164  &  0.7694 \\
         PIMD&32 CPUs& 0.209343&3.4891\\ 
    \end{tabular}
    \caption{Performance comparison of the different MD methods employed to generate the phonon dispersions shown in Fig. \ref{fig:neon_lj_15k_ff} and Fig. \ref{fig:neon_lj_15k_xx}. The second column indicates the computational resources utilized for each approach. The wall time required to perform a single MD step is reported in the third column, while the total wall time required to generate a trajectory of length 60 ps is reported in the fourth column.}
    \label{tab:performance_comparison_neon}
\end{table}

\subsection{Temperature-dependent phonon frequency shifts}
\label{sec:temp_scans}
\begin{figure}[H]
  \includegraphics[width=\linewidth]{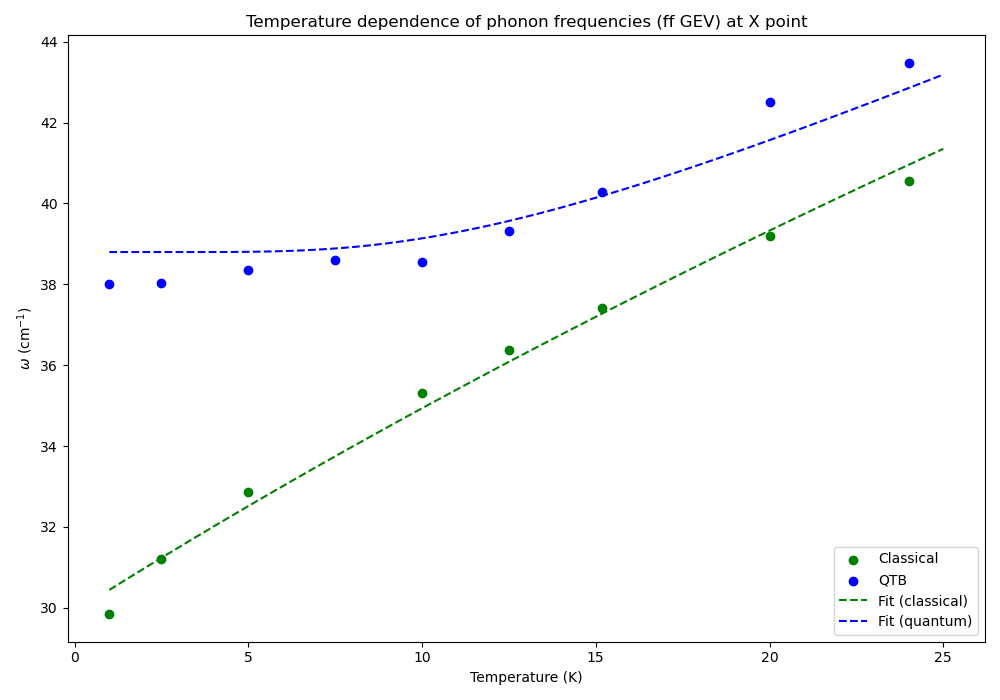}
\caption[Temperature dependence of the frequencies of the transverse acoustic phonons of Lennard-Jones neon at the X point of the Brillouin zone obtained from classical MD and QTB simulations.]{Temperature dependence of the frequencies of the transverse acoustic phonons of Lennard-Jones neon at the X point of the Brillouin zone, obtained from classical MD (green markers) and QTB simulations (blue markers). Fits of the data to the functional form defined by \eqnref{eq:quartic_model} (using the appropriate estimator for $\langle x^2 \rangle$ in the classical and quantum cases respectively) are given by the corresponding dashed lines. The results are presented for the FF-pp GEV.}
\label{fig:neon_temp_scan}
\end{figure}

We performed classical MD and QTB simulations of the dispersions over a range of temperatures from 1K to 24K (which is just below the melting temperature of neon). In all calculations, the volume of the box was fixed to that which was employed in the simulations described in the previous section. Hence, the effect of temperature on the phonon frequencies was studied in the absence of any thermal expansion, as is often reported in the literature \cite{ying2025highly, copley1974measurement, lanzillo2013temperature}. The results of these simulations for the transverse acoustic phonons at the X point of the Brillouin zone are presented in Fig.\ref{fig:neon_temp_scan}. The figure shows noticeable differences between the classical (green dots) and quantum (blue dots) frequencies. The QTB results show an essentially constant value up to about 10K, followed by an increasing trend. In contrast, the classical results show a monotonic increase with the temperature. These trends can be understood from a phenomenological standpoint, and from our framework to determine the phonon frequency. The blue-shift at higher temperatures (for all results) is consistent with atoms penetrating to a larger degree the repulsive part of the potential when the temperature increases. Taking into account the symmetries of the FCC crystal, this can be modelled with a quartic potential along the phonon displacement $x$: $V(x)=\frac{k_0}{2}x^2+\frac{k_1}{12}x^4$. Recalling that the Kubo FF correlation is an estimator of the Hessian multiplied by $k_BT$, and the Kubo pp correlation is an estimator of $mk_BT$, the FF-pp GEV for this model reads
\begin{equation}
\label{eq:quartic_model}
    k_0+k_1 \langle x^2 \rangle  = m \omega^2.
\end{equation}
Classically, the variance $ \langle x^2 \rangle $ is proportional to the temperature $T$: $\langle x^2 \rangle = \frac{k_BT}{k_0}$ to zero order in the anharmonicity. This leads to a square root dependence of the frequency on temperature, which is consistent with the fit shown in the figure as the dashed green curve. Note that the fit parameters lead to a frequency of $\omega^*=29.63$ \cminv{} for $T=0$~K, in good agreement with the previously reported~\cite{efremkin2022study} harmonic value of $\sim 30$ \cminv. In the quantum case at low temperature, on the other hand, the presence of zero point energy implies that $\langle x^2 \rangle$ is bounded below. Indeed, a quantum mechanical treatment yields, to lowest order, $\langle x^2 \rangle = \frac{\hbar}{2 m \omega_0} \coth\left(\frac{\hbar \omega_0}{2 k_B T}\right)$, with $\omega_0=\sqrt{k_0/m}$ being the frequency obtained via a harmonic approximation of the potential. Thus, in the quantum case, one has $\langle x^2 \rangle \to \frac{\hbar}{2 m \omega_0}$ as $T \to 0K$. This explains the saturation of the quantum phonon frequency to a value higher than that of the classical case. The extrapolated value of the quantum frequency at zero temperature is 38.8~\cminv{} in good agreement with the computed value at T=1K, as can be seen from the figure. Contrariwise, in the high temperature limit $\langle x^2 \rangle \to \frac{k_B T}{k_0}$, and therefore, as observed in \figref{fig:neon_temp_scan}, the quantum phonon frequency starts to follow a $T^{1/2}$ trend, similarly to the classical case. %As a consequence, the quantum frequencies remain largely constant up to about 10K. Only when the temperature has been raised to around 10K do thermal kinetic effects lead to an increase of the variance  of the repulsive region and thus further entering the repulsive region and recovering an increase of the phonon frequencies when temperature is further increased. \tb{I'm not sure this last sentence is super clear. How about "Only when the temperature has been raised to around 10K do thermal kinetic effects lead to a sufficient increase in the variance, $\langle x^2 \rangle$, such that further penetration of the repulsive region is facilitated, thus inducing a further increase in the phonon frequencies."}

\section{Conclusions}
\label{sec:conclusion}
In this work, we have outlined a strategy to compute phonon dispersions when anharmonic and nuclear quantum effects are relevant. The method is accurate and requires relatively low numerical effort. The approach combines QTB trajectories (to account for quantum effects) and the quantum correlators method (to obtain phonon frequencies and eigenmodes). The validity of this novel combination of methodologies was first investigated by studying one-dimensional toy models and then phonons in a Lennard-Jones crystal of FCC neon. Owing to the relative simplicity of the description of the interatomic forces in this system, we were able to perform a detailed comparison of the phonon dispersions obtained using QTB and reference PIMD results. The agreement between the two sets of results was found to be very good, with QTB clearly capturing the effects of quantum nuclear motion on the vibrational properties of the system. These investigations demonstrate, in particular, that results derived from the force-momentum GEV are reliable and straightforward estimators of the dispersions. Use of the coordinate-velocity GEV is more delicate due to the presence of spurious low frequency signals that require ad hoc corrections. 

The low numerical cost of our approach makes it particularly suitable to investigate properties that require multiple evaluations of the phonon dispersions, like the study of the temperature dependent frequency shifts presented in this work. Leveraging QTB trajectories will also facilitate the extension of the size of phonon calculations based on an ab initio electronic structure description of the system and investigation of the more quantum regimes where PIMD approaches become impractical.

\begin{acknowledgments}
The authors are grateful to S. Huppert, T. Pl\'e and T. Morresi for insightful discussions. 

This research was supported by the NCCR MARVEL, a National Centre for Competence in Research, funded by the Swiss National Science Foundation (grant number 205602). 
\end{acknowledgments}

The authors have no conflicts to disclose.

\section*{Data Availability Statement}
Data available on request from the authors

\appendix

\section{Details of QTB calculations}
\label{appendix:QTBadQTB} 
In this Appendix, we discuss the most relevant aspects of the QTB implementation deployed in this work. For the sake of keeping the discussion self-contained, when need be we provide a brief summary of methods detailed elsewhere.

A critical step of the algorithm is generating the colored noise described in \eqnref{qtb_fdt} and in the following text. Here, this is done on-the-fly, using the scheme prescribed in Ref.\citen{barrat2011portable}. Firstly, Gaussian white noise $\xi(t)$ is created in the time domain, with zero mean and unit variance. This noise is then transformed into the frequency domain using a Fourier transform, yielding $\tilde{\xi}(\omega)$. Once in the frequency domain, the noise is filtered by multiplying it with the square root of the quantum spectral density and a cutoff function, ensuring that unphysical high-frequency noise is suppressed. The modified noise in frequency space is given by

\begin{equation}
    \tilde{F}(\omega) = \sqrt{C_{FF}(\omega)} \tilde{\xi}(\omega) \cdot g(\omega),
\end{equation}
where $g(\omega)$ is the cutoff function. In our calculations:

\begin{equation}
    g(\omega) =   1+ \exp((\omega-\omega_c)/\omega_{s})
\end{equation}

This cutoff function ensures that noise at frequencies higher than $\omega_c$ is gradually suppressed (with the smoothness of the cutoff determined by the smearing parameter $\omega_s$ which is typically chosen to be $\sim \frac{\omega_c}{50}$), preventing inclusion of unphysically high frequencies into the dynamics. The filtered noise is then transformed back into the time domain using an inverse Fourier transform, yielding the final time-dependent stochastic force $F(t)$. This force is subsequently applied to the system within the Langevin dynamics. The choice of the cutoff frequency itself is largely governed by the maximum physical frequency present in the system under consideration, which we shall refer to as $\omega_0$. For a crystalline solid for instance, a good estimate for $\omega_c$ is a few multiples of the Debye frequency, $\Theta_D$, of that solid, with $\Theta_D$ roughly corresponding to this maximum frequency. The value of the cutoff adopted for neon is reported in the main text.

The other primary concern that arises when setting up a QTB simulation is the choice of the damping coefficient, $\gamma$, employed in the thermostat. Prior analyses \cite{barrat2011portable} have suggested that an appropriate range of $\gamma$ is typically such that the ratio $\gamma/\omega_0 \sim 10^{-2}$. There are two competing considerations that must be taken into account when making this choice. On one hand, larger values of $\gamma$ more effectively suppress zero-point energy leakage (ZPEL) in the system. On the other hand, minimizing the value of $\gamma$ avoids undesirable broadening of the QTB response function. In practice, one should weigh these two factors against one another in short trial runs to determine a suitable value for $\gamma$. An additional consistency check on these choices of $\omega_c$ and $\gamma$ is to carry out a short PIMD run and use the converged value of the centroid-virial estimator of the PIMD kinetic energy as a reference against which the QTB kinetic energy can be compared. Inaccurate choices of $\omega_c$ and, to a lesser degree, $\gamma$, shall manifest as noticeable over- or underestimations of the kinetic energy with respect to the PIMD reference.

\begin{figure}[H]
    \centering
    \includegraphics[width=1.0\linewidth]{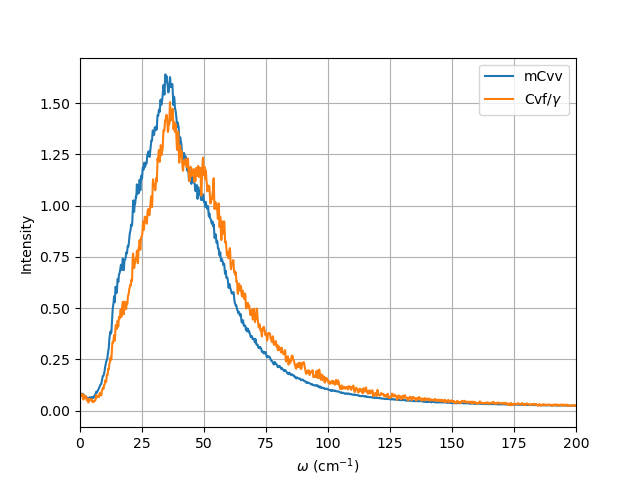}
    \caption{Plot of the two sides of the fluctuation-dissipation relation (\eqnref{eq:qtb_fdr}) for our standard QTB simulations of Lennard-Jones neon at 15.185K and using a friction coefficient of $\gamma=2\times 10^{-5}$ a.u.}
    \label{fig:neon_fdr}
\end{figure}

\begin{figure}[H]
    \centering
    \includegraphics[width=1.0\linewidth]{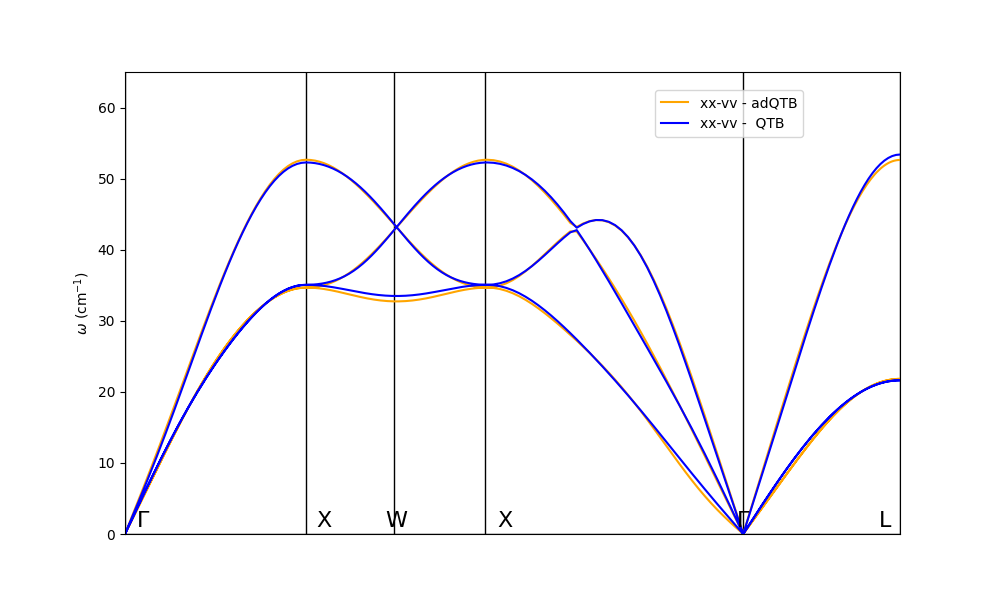}
    \caption[Phonon dispersions for Lennard-Jones neon at 15.185K using  QTB (blue curves) and adQTB (orange curves).]{Phonon dispersions for Lennard-Jones neon at 15.185K using QTB (blue curves) and adQTB (orange curves). Kubo-transformed TCFs have been used in constructing the correlators in the xx-vv GEV.}
    \label{fig:neon_lj_15k_adqtb_xx}
\end{figure}

\begin{figure}[H]
    \centering
    \includegraphics[width=1.0\linewidth]{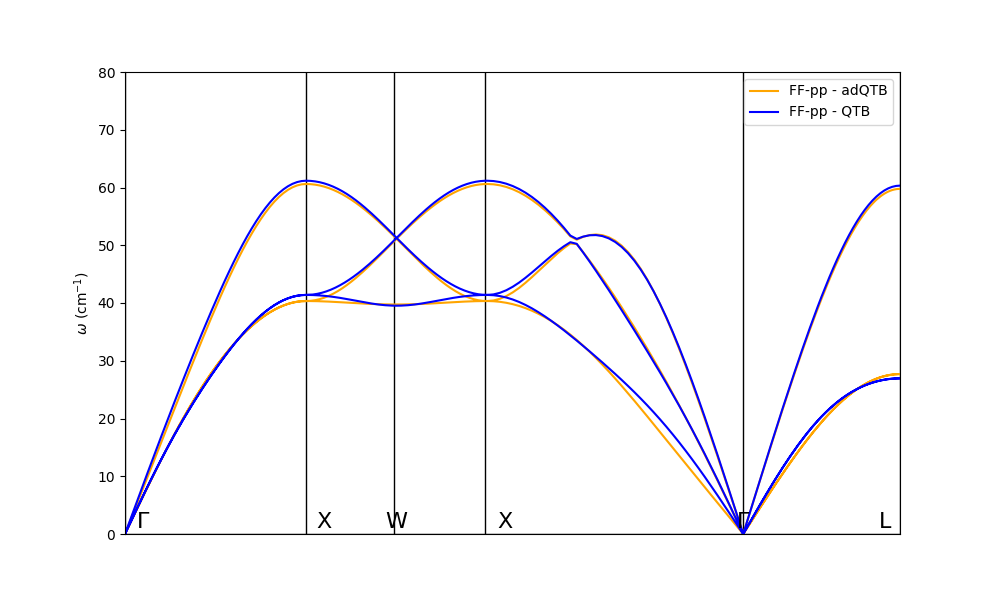}
    \caption[Phonon dispersions for Lennard-Jones neon at 15.185K using  QTB (blue curves) and adQTB (orange curves).]{Phonon dispersions for Lennard-Jones neon at 15.185K using QTB (blue curves) and adQTB (orange curves). Kubo-transformed TCFs have been used in constructing the correlators in the FF-pp GEV.}
    \label{fig:neon_lj_15k_adqtb_ff}
\end{figure}

As discussed in the main text, ZPEL can affect the accuracy of QTB calculations and the so-called adaptive quantum thermal bath (adQTB) scheme has been proposed to mitigate this pathology.  adQTB monitors the level of ZPEL present in the system by evaluating the degree to which the quantum fluctuation-dissipation relation is violated. A quantitative measure of this violation is given by the following metric, which can be computed on-the-fly in a QTB simulation: 
\begin{equation}
\label{eq:qtb_fdr}
  \Delta_{\text{FDT},i}(\omega) = \text{Re}[C_{vF,i}(\omega)] - m_i\gamma_i(\omega) C_{vv,i}(\omega)
\end{equation}. Here, $C_{vF,i}(\omega)$ and $C_{vv,i}(\omega)$ are, respectively, the random force-velocity and velocity-velocity autocorrelation functions of the i-th Cartesian degree of freedom.  Correction of the fluctuation-dissipation relation is achieved, at least in one variant of the technique, by adapting the friction parameter of the bath, $\gamma$, in a mode-dependent manner (i.e., so that $\gamma=\gamma(\omega)$) in order to minimize the absolute value of $\Delta_{\text{FDT},i}(\omega)$ for each value of $\omega$ and each degree of freedom $i$. 
To investigate the relevance of ZPEL in the calculations reported in the main text for neon, we monitored ~\eqnref{eq:qtb_fdr} in preliminary runs. Fig.\ref{fig:neon_fdr} indicates that although there is a very small degree of leakage, the FDR is in general well-satisfied for our system. The agreement could be improved by further increasing the damping coefficient of the  thermostat, but this is found to adversely affect the phonon dispersions as a consequence of overestimation of the system's kinetic energy. Indeed, in our experience, it is important to strike a balance, taking into account these different factors when choosing the value of $\gamma$. The very limited impact of adaptation on the calculations presented in this work is confirmed by examining the phonon dispersions. Fig.\ref{fig:neon_lj_15k_adqtb_xx}, clearly shows that adaptation does not significantly impact the phonon dispersions in the case of the xx-vv GEV and so our use of standard QTB is warranted. These remarks are also largely true for the FF-pp GEV, see Fig.~\ref{fig:neon_lj_15k_adqtb_ff}, where small differences between the brute force and adapted QTB can only be seen in a small range of wavevectors along the X-$\Gamma$ segment. The dispersions for the adQTB runs shown in the figures were obtained with an initial value of $\gamma=2\times10^{-5}$ a.u.

\bibliographystyle{unsrt} 
\bibliography{refs} % Entries are in the refs.bib file

\end{document}